\documentclass[%
 amsmath,amssymb,
 aps,
prfluids
]{revtex4-2}

\usepackage{graphicx}
\usepackage{dcolumn}
\usepackage{bm}
\usepackage{hyperref}

\usepackage[
text={7in,10in},centering,
]{geometry}

\usepackage{lineno}

\begin{document}

\title{Measuring the viscoelastic behavior of dilute polymer solutions using high-speed statistical particle microrheology}


\author{Zijie Qu}
 \altaffiliation[Current address: ]{California Institute of Technology, The Division of Biology and Biological Engineering, 1200 East California Boulevard, Pasadena, California, 91125 USA.\\
 Email: zijiequ@caltech.edu}
 
 \author{Xiongfeng Yi}
 \altaffiliation[Current address: ]{4726 Calhoun Road, Houston TX 77204.\\
 Email: Xyi6@uh.edu}
 
\author{Kenneth S. Breuer}%
\affiliation{%
 Brown University, School of Engineering, 184 Hope St, Providence, RI 02912 USA
}%

\date{\today}

\begin{abstract}
The viscoelastic behavior of polymer solutions is commonly measured using oscillating shear rheo\-metry, however, the accuracy of such methods is limited by the oscillating frequency of the equipment and since the relaxation time of the dilute polymer solutions is short, this requires measurement at very high frequencies. Microrheology has been proposed to overcome this technical challenge. Yet the equipment for resolving the statistics of particle displacements in microrheology is expensive. In this work, we measured the viscoelastic behavior of Methocel solutions at various concentrations using a conventional epi-fluorescence microscope coupled to a high-speed intensified camera. Statistical Particle Tracking is used in analyzing the mean-squared displacement of the dispersive particles. Relaxation times ranging from 0.76 - 9.00 ms and viscoelastic moduli, $G'$ between 11.34 and 3.39 are reported for Methocel solutions of concentrations between 0.063 - 0.5\%.
\end{abstract}
\maketitle

\section{Introduction}
\label{intro}
The rheology of complex fluids has been studied extensively through the last few decades \cite{han1976rheology,baumgartel1996relaxation,bird1977dynamics}. It is highly related to not only industrial applications, such as paints \cite{hester1997rheology}, plastics \cite{karato1995plastic} and printing inks \cite{zettlemoyer1955rheology}, but also to many research areas especially on biological application and processing \cite{amblard1996magnetic,cokelet1963rheology}. The rheological properties, especially the viscoelasticity of complex fluids, give important details on micro-structural features and dynamics of the system \cite{ferry1980viscoelastic}. In general, the relaxation time of the system varies and spans over a large range of time scale depending on this viscoelastic modulus ($G'$ and $G''$) \cite{crassous2005characterization,squires2010fluid}. For dilute polymer solutions with small molecule size, the relaxation time is usually short. Measurements with high frequency are indispensable to resolve their rheological behavior \cite{mason1997particle,fritz2003characterizing}.

The viscoelastic behavior of such solutions can be measured using an oscillating shear rheometer \cite{arbogast1997high,willenbacher2007dynamics}, which however requires a precise and sensitive measurement of the torque of the shear plate at a high frequency and is difficult to achieve using a conventional rheometer \cite{willenbacher2007dynamics}. Microrheology based on the particle dispersion in non-Newtonian solutions has been previously proposed \cite{mason1995optical,mason1997particle} to overcome such difficulties. The typical experimental technique in microrheology for resolving the statistics of the particle displacement is Dynamic Light Scattering (DLS) \cite{mason1995optical}. However, the experimental system requires specialized equipment \cite{berne2000dynamic} which makes it hard and expensive to implement. 

The most general way of measuring particle displacements or mean-squared displacements (MSD), which are required for microrheology \cite{squires2010fluid}, is to compare the changes in the position of the same particle between consecutive images. However, the positions of small fluorescent particles are difficult to detect accurately with short exposure at a high frame rate. Camera noise, especially generated from CMOS cameras, including the general Gaussian noise \cite{cattin2013image}, photon shot noise \cite{macdonald2006digital} and signal read noise becomes more severe as the image acquisition frequency increases. An image intensifier is capable of detecting and amplifying low-light-level images to overcome the limited exposure at high frame rate for fluorescent systems \cite{qian2015large}. However the intensifier brings additional noise to the resultant image. All these noisy signals may lead to a false measurement of particle displacements during image post-processing. In addition to optical noise, overlapping of particles due to high dispersion can also bring difficulties in displacement detection \cite{guasto2006statistical}.

Another drawback of traditional particle displacement detection is that it requires a precise particle-to-particle matching between frames. The most widely used algorithm for particle matching is nearest-neighbor matching \cite{schmidt1996imaging}, which however does not guarantee a correct match and becomes less accurate with all the false signals introduced at high frequency. Even in dilute particle suspensions, it is possible that multiple particles cluster locally and the nearest-neighbor algorithm does not lead to a one-to-one matching. Particles may also disperse out of focus or outside the observation window that a particle pair do not physically exist. For images taken at high frequency, some background noise can be detected as ''particles" and such random appearance breaks the one-to-one matching using the nearest neighbor algorithm. 

To address these issues when measuring the MSD at high frequency, Statistical Particle Tracking Velocimetry (SPTV) technique \cite{guasto2006statistical} is applied to the image and data post-processing. With a similar approach, SPTV requires all particles in each frame to be detected using intensity threshold \cite{levi20053} or diffraction ring \cite{afik2015robust}. Instead of finding a precise particle matching, SPTV purposely utilizes a large interrogation window to include multiple tracer particles (and noise ``particles") and measures the displacement distribution. Drop in/drop out particles due to dispersion and noise ``particles" are all included for particle matching to generate the displacement statistics, but are eliminated in post-processing by exploiting the fact that true particle-particle motion will observe a diffusive (Gaussian, or near-Gaussian) displacement distribution, while spurious particle-particle or particle-noise tracking will observe a uniform distribution. As long as the number of particles tracked is large, and as long as the medium is uniform and isotropic, the width of the measured particle distribution function can be directly related to the MSD of a particle in the fluid \cite{guasto2006statistical}.  

In this paper, we use SPTV, as suggested by Guasto \emph{et al.} \cite{guasto2006statistical}, to overcome the difficulties in implementing traditional particle displacement measuring techniques.  We present a microrheological assessment of the viscoelastic behavior of dilute concentrations of Methocel 90 HG - a long chain polymer that exhibits viscoelastic properties and low shear viscosities (up to $18$ cP).

\section{Experimental setup and procedure}
\begin{figure}
\centering
\includegraphics[width=0.5\columnwidth]{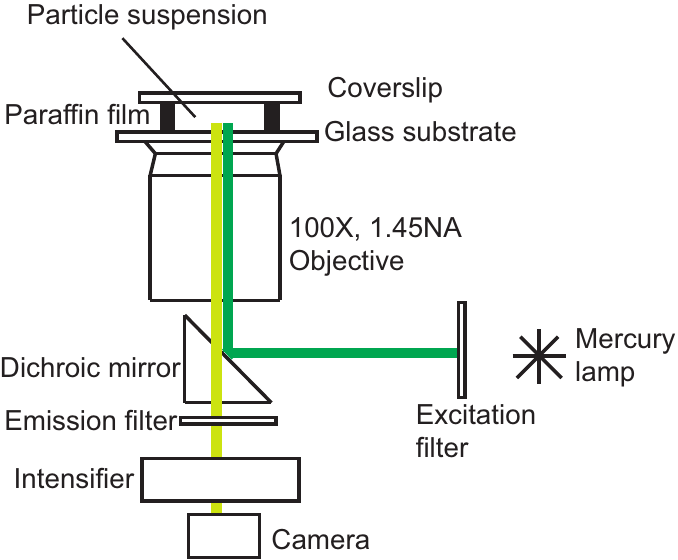}
\caption{Schematic of the experimental configuration. The tracer particles are illuminated using a mercury lamp with an excitation filter, the beam is introduced along the optical axis. The emission light follows the same light path until it reaches the dichroic mirror and passes through the emission filter. The light is captured via an image intensifier by a CMOS camera at up to 1000 fps.}
\label{fig:fig1}
\end{figure}

The schematic of the experimental configuration is shown in Fig.~\ref{fig:fig1}. A $0.500$\% (wt/vol)  Hydroxypropyl Methyl-cellulose (Methocel 90 HG, Sigma Aldridge) stock solution was prepared by dissolving the polymer in deionized water and rotating overnight at $200$ rpm. Lower concentration solutions were diluted from the stock solution. The particle suspensions were prepared by diluting 200 nm and 100 nm fluorescent beads (Molecular Probes, Excitation/Emission: 540/560nm) 500,000 times into each polymer solution. 

A simple test fixture was fabricated using a small piece of Paraffin film (Parafilm M), cut into a square ($\sim$ $1$ cm) with a small hole ($\sim$ $0.3$ cm) punched through in the middle to create a test well. The film was placed upon a No.1 coverglass (Fisher Scientific) and heated gently for $30$ sec. until the film started to soften. A small volume ($50$ $\mu$L) of the test fluid was placed in the test well and sealed with a No.1.5 coverslip (Fisher Scientific) on top. The sample was allowed to cool to room temperature for $5$ min before the experiment commenced. 

The motion of the particles was observed using an inverted epifluorescent microscope (Nikon ECLIPSE TE 2000 - U) equipped with a 100X oil immersion objective (Nikon, Plan Apo TIRF). Images were recorded using a high-speed CMOS camera (Photron Fastcam SA - 5) fitted with a high-speed image intensifier (Hamamatsu V9501U - 74 - G240).

To resolve the frequency-dependent viscoelastic moduli of the polymer solutions at various concentrations. Images were recorded at different framerates \cite{mason1995optical} ranging from $5$ Hz to $1000$ Hz using Photron FASTCAM Viewer (PFV) software. $2000$ images were saved for each sample suspension and each experiment was repeated $3$ times to reduce measurement errors and to ensure consistency.

\section{Image analysis and data processing}
Images were processed using custom software written in C++ and using the OpenCV library. A bilateral filter \cite{zhang2008adaptive} was applied to the image to reduce background noise. This filter was used since it reduces the noise without sacrificing the sharp intensity gradient near edges of the imaged particles \cite{jiang2003applications}. Following the application of the filter,  the position of the centroid of each particle (including phantom particles due to background noise) was detected using the $contour$ and $moment$ functions in the OpenCV library. 

To compute particle displacement distributions for a given time separation, $\Delta t$, the distance between all particles in the first image and all particles in the second image was computed. Unlike traditional particle tracking algorithms, there was no attempt to find likely particle pairs between the two images.  This procedure was repeated for all image pairs separated by $\Delta t$, generating a distribution of particle displacements for each value of $\Delta t$.  The distribution thus included all physical particle displacements, as well as ``displacements'' associated with random particle pairings as well as particle-noise and noise-noise pairings (see \cite{guasto2006statistical} for more details).

As suggested by Guasto \emph{et al.} \cite{guasto2006statistical}, the distribution of particle displacements in any direction  $R(t)$ should have the form of  a modified Gaussian distribution with zero mean:
\begin{equation}
S(R, t) = C(t) + \frac{1}{\sqrt{2\pi\sigma(t)}} \exp (-\frac{(R -\mu(t))^2}{2\sigma(t)^{2}}),
\label{equation1}
\end{equation}
which is simply a Gaussian distribution plus a constant offset, or ``table''. The Gaussian part is due to the correlated particle displacements \cite{mason2000estimating} while the table derives from phantom displacements recorded between un-correlated particle pairs and particle-noise pairs \cite{guasto2006statistical}. Since these parings are random the table has a uniform distribution. Note the mean-squared displacement (MSD) of particle dispersion is defined as
\begin{equation}
MSD(t) = <( R(t) - R(0))^2> = <R(t)^{2}>,
\end{equation}
and by definition, the standard deviation $\sigma(t)$ of the Gaussian part in Equation.~\ref{equation1} is given as
\begin{equation}
\sigma(t) = \sqrt{<(R (t) - \mu(t))^2>}.
\end{equation}
$\mu(t)$ represent the mean speed, or drift velocity of the flow field, which in the present case is close to $0$. $\sigma(t)^{2}$ is then a good estimation of $MSD(t)$.

The measured displacements histogram is fitted to Equation.~\ref{equation1} (using MATLAB's $fitnlm$ function) to retrieve the $MSD$. The size of the interrogation window ($l$) is chosen to be larger than at least three times $\sigma_{i}(t)$ for each fitting process, otherwise the $MSD$ is often overestimated.  

\subsection{Viscoelastic spectrum calculation}
To accurately compute the relaxation time ($\tau$), we follow the method proposed by Mason \emph{et al.} \cite{mason1997particle} which is repeated here for convenience. Assuming the fluid is isotropic and incompressible, the fluids viscoelastic spectrum $\widetilde{G} (s)$ is calculated as
\begin{equation}
\widetilde{G} (s) \approx \frac{k_{B}T}{\pi r MSD(t) \Gamma [1+(\partial ln(MSD(t))/ \partial ln(t))]} \Big|_{t=1/s}
\end{equation}
where $k_B$ is the Boltzmann constant, $T$ is the absolute temperature, $r$ is the radius of the particle and $\Gamma$ stands for the Gamma-function. We calculate the fluids viscoelastic spectrum $\widetilde{G}(s)$ using the result from measured MSD over time. The partial derivative is numerically approximated using a first-order finite difference. 

A non-linear curve fit of the resulting $\widetilde{G} (s)$ is performed to find the relaxation time(s), $\tau_{i}$, and constant(s), $G_i$:
\begin{equation}
\widetilde{G} (s) = \sum_{i} \frac{G_{i}s}{s+1/\tau_{i}}.
\end{equation}

The storage ($G'$) and loss ($G''$) modului for polymer solutions at different concentration  for each relaxation time are calculated as:
\begin{equation}
G' (\Omega) = \sum_{i} \frac{G_{i} \tau_{i}^{2} \Omega^{2}}{1+\Omega^{2} \tau_{i}^{2}},
\end{equation}
and
\begin{equation}
G'' (\Omega) = \sum_{i} \frac{G_{i} \tau_{i} \Omega}{1+\Omega^{2} \tau_{i}^{2}}.
\end{equation}

\section{Results and discussion}
\begin{figure}[h]
\centering
\includegraphics[width=0.6\linewidth]{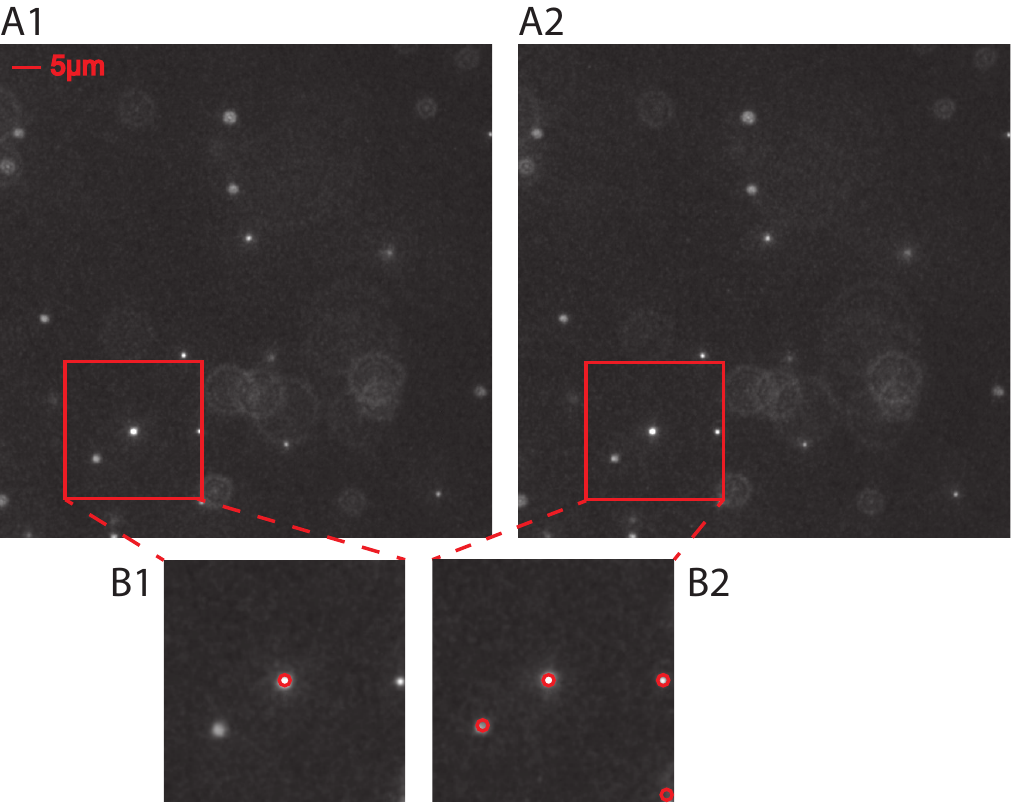}
\caption{Particles ($200$ nm) dispersion in  $0.125 \%$ Methocel solution at two consecutive frames (A1 and A2). B1 and B2 are zoom-in images of the parts shown in A1 and A2 by a square (interrogation window) respectively. Circle marker in B1 shows the central particle and markers in B2 are all particles detected within the same interrogation window in the following image. The displacements are measured between all particles in B2 to the one in B1.}
\label{fig:fig2}
\end{figure}

Sample images are shown in Fig.~\ref{fig:fig2} illustrating the seeding particles (diameter: $200$ $nm$) in $0.125 \%$ Methocel solution.  Fig.~\ref{fig:fig2} A1 and A2 are two consecutive images taken at $250$ fps, while images B1 and B2 are zoomed-in parts of each image denoted by the red squares, illustrating the interrogation window. 

\begin{figure}[h]
\centering
\includegraphics[width=0.5\columnwidth]{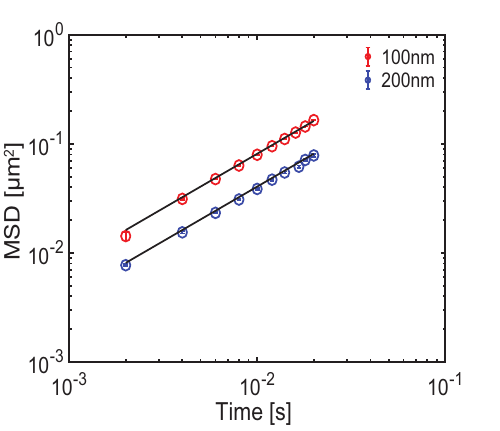}
\caption{MSD of diffusive particles in water. Particles with two sizes ($100$ nm and $200$ nm) are tested to calibrate the measurement and processing algorithm. The black lines are estimated MSD over time theoretically using Stoke-Einstein equation. Temperature is $18^{\circ}$C. }
\label{fig:fig3}
\end{figure}

The procedure was first validated using water, a Newtonian fluid which should exhibit an MSD growing linearly with time.  The displacement distributions of $100$ and $200$ nm particles were measured with $\Delta t$ varying from 0.002 - 0.02 seconds. The MSD was computed using the SPTV technique described above, and the results are showing in Fig.~\ref{fig:fig3}. The slope of the measured MSD is in good agreement with the theoretical approximation based on the Stokes-Einstein equation \cite{einstein1956investigations} denoted by the solid lines.

\begin{figure}[h]
\centering
\includegraphics[width=0.5\columnwidth]{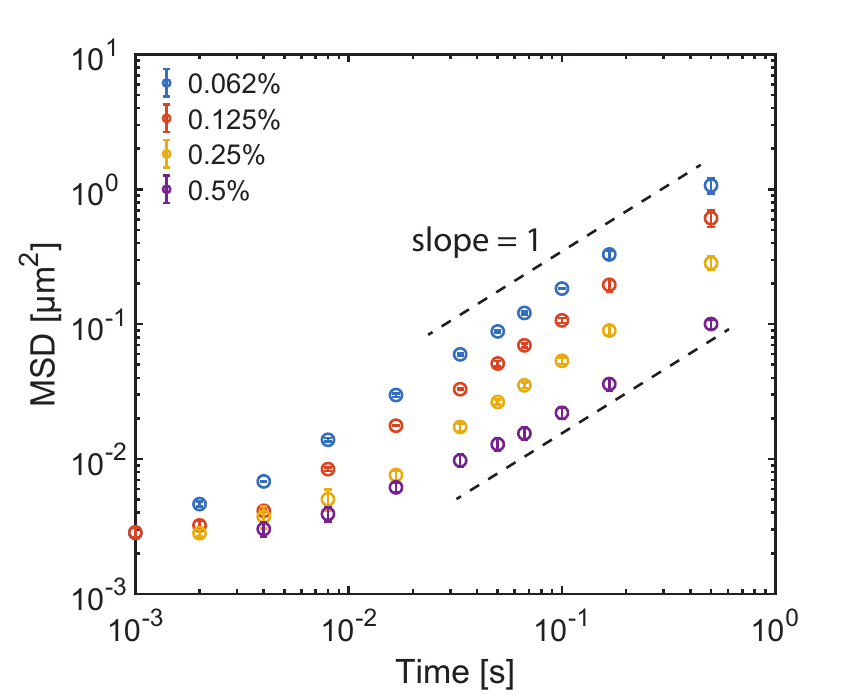}
\caption{MSD of particle dispersion in Methocel solutions at different concentrations as a function of time. The slope of MSD over time is $1$ in four different concentrations at long time scale and becomes less than $1$ at short time scale. In addition, the transition time increases with respect to polymer concentration.}
\label{fig:fig4}
\end{figure}

With this reassurance, the MSDs of $200$ nm particles suspended in Methocel solutions of different concentrations were measured and are shown in Fig.~\ref{fig:fig4}. It is observed the MSD decreases as the polymer concentration increases indicating an increased viscous behavior. For long times the slope of MSD (on a log-log scale) is $1$ for all concentrations used in the experiment, suggesting a diffusive behavior. At shorter times, however, the slope becomes less than $1$ indicating the existence of elastic behavior in the fluid \cite{mason1997particle}. Moreover, it is directly observed from Fig.~\ref{fig:fig4} that the crossover time (relaxation time), which is indicated by the transition point when the slope of the curve becomes less than $1$ increases with respect to polymer concentration.

\begin{figure}[h]
\centering
\includegraphics[width=0.8\columnwidth]{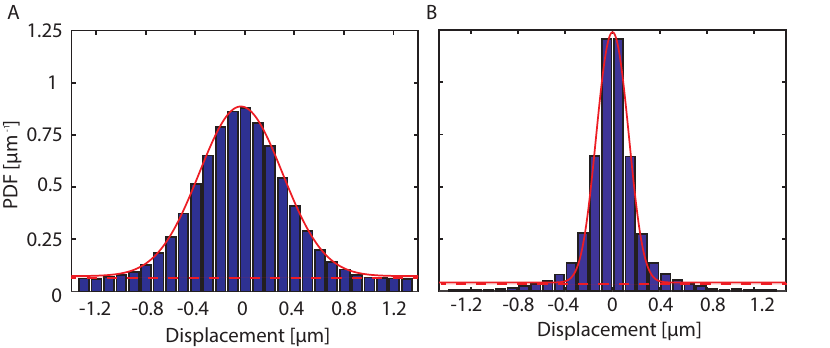}
\caption{Measured particle displacement distribution. A, the displacement distribution of 100 $nm$ particle in water taken at 100 Hz. B, the displacement distribution of 200 $nm$ particle in 0.125\% Methocel solution take at 125 Hz. The solid red curves are the fitted Gaussian distribution and the dashed red lines indicate the constants used in each case. }
\label{fig:fig5}
\end{figure}

It has been observed previously that for particle dispersion in non-Newtonian fluids, the shape of the displacement distribution  has a Gaussian center, but a non-Gaussian tail. This is particularly true when the time separation is close to polymer relaxation time \cite{weeks2000three,toyota2011non}. However, the non-Gaussian tail contributes only $5 \%$ to the total distribution as suggested by Weeks \emph{et la.} \cite{weeks2000three}, which introduces a subtle change to the estimated MSD as compared with the one found by assuming a pure Gaussian distribution. This effect is observed in this study, resulting in the overestimation of the distribution tail  (Fig~\ref{fig:fig5} B). However, since the key information required for the MSD is contained in the width of the distribution, and not the tail, the discrepancy is negligible, and we are able to ignore it in our analysis.

\begin{figure}[h]
\centering
\includegraphics[width=0.8\linewidth]{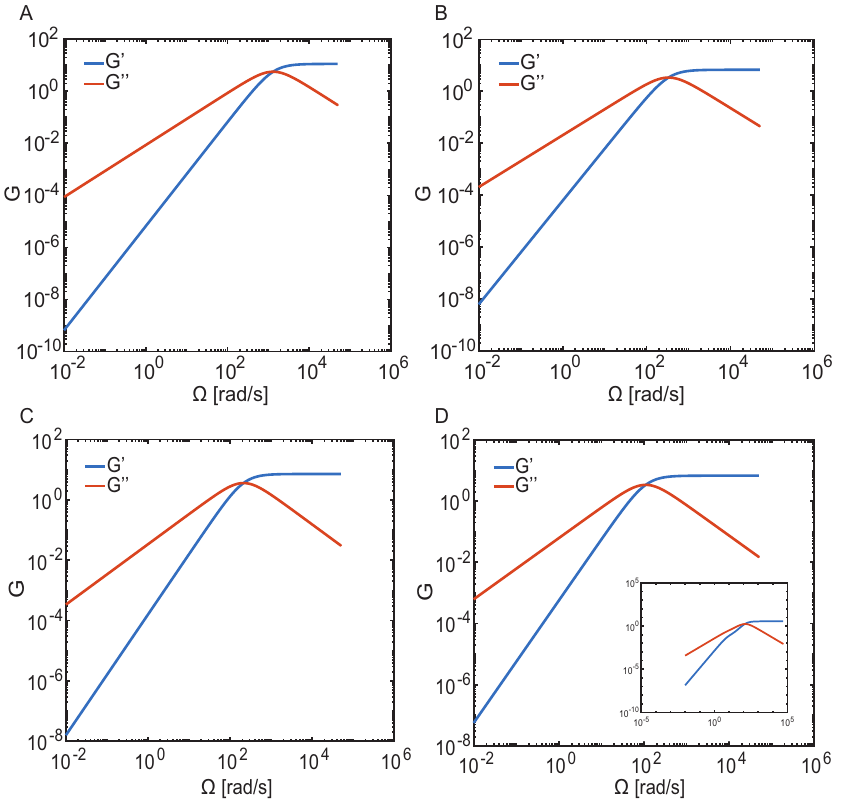}
\caption{Calculated elastic and viscous modulus of the Methocel solutions tested in the experiment at different concentration. A, 0.063\% B, 0.125\% C, 0.250\% D, 0.500\%. Inset D, 0.500\% with two relaxation times. The non-linear fitting model assumes a single relaxation time as it is similar to the result of multiple relaxation time. The cross over point, which also indicates the relaxation time increases as a function of polymer concentration which is observed from Fig.~\ref{fig:fig4}.}
\label{fig:fig6}
\end{figure}

The relaxation time $\tau_{i}$ and viscoelastic moduli $G'$, $G''$ were calculated, as described above. The results are shown in Fig.~\ref{fig:fig6} and Table 1. The relaxation time of the polymer solutions we tested increases as the concentration rises, which is indicated by the cross-over frequency of $G'$ and $G''$ in Fig.~\ref{fig:fig6}. The calculation was carried out assuming both one and two relaxation times, although adding a second pair of fitting parameters ($\tau_2, G_2$) had a negligible effect on the estimation of the longest time, $\tau_1$ and modulus, $G_1$ (Table 1). This is also shown in the inset figure in Fig.~\ref{fig:fig6} D. 

\begin{table}[h]
\centering
\caption{Relaxation times for different Methocel concentrations, assuming one and two relaxation times ($i = 1,2$).}
\begin{tabular}{llrll} 
\multicolumn{3}{l}{$i = 1$}\\
Conc. & $\tau_1$ [ms] & $G_1$  \\ 
\hline     
0.063 &      0.76  & 11.34 & \\
0.125 &      2.99  & 6.76 & \\ 
0.250 &      4.68  & 7.23 & \\
0.500 &      9.00  & 3.39 & \\
\hline
\multicolumn{5}{l}{$i = 2$}\\ 
Conc. & $\tau_1$ [ms] & $G_1$  & $\tau_2$ [ms] & $G_2$  \\ 
\hline
0.063 &      0.76  & 11.34  & 1.00$\times 10^{-9}$     & 1.00$\times 10^{-3}$\\
0.125 &      3.13  & 6.65   & 4.53$\times 10^{-6}$     & 3.13$\times 10^{-3}$ \\ 
0.250 &      4.70  & 7.20   & 2.42$\times 10^{-5}$     & 1.92$\times 10^{-2}$ \\
0.500 &      8.99  & 3.46   & 5.43$\times 10^{-4}$     & 6.52$\times 10^{-6}$\\
\hline
\end{tabular}

\label{table:1}
\end{table}

\section{Concluding remarks}
Statistical Particle Tracking Velocimetry (SPTV)  has been introduced and applied to a microrheology experiment as a replacement of more conventional tracking techniques. The advantages of SPTV are that it removes concerns of spurious particle tracking which is particularly problematic when tracking small particles with highly-amplified intensified images.  The data presented here demonstrate that the sampling frequency of microrheology measurement with intensified high speed image acquisition and fluorescent microscopy works up to $1$ kHz. It thus extends the measured spectrum of a conventional mechanical shear rheometer. The result indeed shows the relaxation time of dilute Methocel solutions, which is on the order to $10$ $ms$, are short. Our result provides a robust basis for a complete study of the viscoelastic behavior of dilute polymer solutions.

Despite its appeal, this method is still limited. First, the rheology of the fluid is resolved at zero shear rate. Second, for solutions with higher viscosity, smaller-sized particles (or quantum dots) are needed for displacement detection \cite{guasto2006statistical}, which may not be guaranteed to observe the Stokes-Einstein relation even in Newtonian solutions \cite{guasto2006statistical}. Lastly, the SPTV method requires a known physical distribution which makes it difficult to implement on particle dispersion in concentrated polymer solutions, where the distribution is non-Gaussian \cite{weeks2000three,toyota2011non}.

\begin{acknowledgements}
This work was supported by the National Science Foundation, Grant CBET 1336638)
\end{acknowledgements}

\bibliography{mainNotes}

\end{document}